\documentstyle[twoside,fleqn]{article}
\newcounter{zacountsec} 
\newcommand{\eh}{\hfill}\newlength{\sperr} 
\newcommand{\Title}[1]{{\large \bf #1}} 
\newcommand{\Author}[1]{{\sc #1}} 
\newcounter{symposium}
\newcommand{\Symposium}[1]{\setcounter{symposium}{#1}} 
\newcounter{session}
\newcommand{\Session}[1]{\setcounter{session}{#1}} 
\expandafter\ifx\csname jtitle \endcsname\relax\def\jtitle{}\fi 
\expandafter\ifx\csname symptitle \endcsname\relax\def\symptitle{}\fi 
\expandafter\ifx\csname makeheadings \endcsname\relax\def\makeheadings{}\fi 
\newcommand{\Section}[1]{{\stepcounter{zacountsec}\vspace{3mm}%
\hspace*{18mm}\normalsize\bf\arabic{zacountsec}. \parbox[t]{150mm}{ #1 }}} 
\newenvironment{Abstract}{\begin{minipage}[t]{177mm}\em }
{\end{minipage}} 
\newenvironment{thm}[2]{\begin{sloppypar}
{#1 #2.}\em{}}
{\end{sloppypar}}
\newcommand{\proof}{\hspace*{9mm}{\settowidth{\sperr}{\rm Proof}%
\parbox[t]{1.3\sperr}{\rm P\eh r\eh o\eh o\eh f\eh. } }}
\newcounter{zalit}
\newenvironment{Acknowledgements}{\vspace{3mm}
\hspace*{18mm}{\bf Acknowledgements}\\[0.3cm]\begin{minipage}[t]{177mm}%
\small \em}{\end{minipage}}
\newenvironment{References}{
\Section{References}
\begin{small}\begin{list}{\arabic{zalit} }{\usecounter{zalit} 
\itemsep0mm \parsep0mm\settowidth{\labelwidth}{\small\rm 88}\labelsep0mm 
\setlength{\leftmargin}{\labelwidth}}}
{\end{list}\end{small}}
\newlength{\addrt}
\newenvironment{Address}{
\begin{minipage}[t]{\addrt}}
{\end{minipage}}
\newcommand*{\AuthorID}[1]{}\newcommand*{\LastName}[1]{}
\newcommand*{\FirstName}[1]{}\newcommand*{\ShortFirstName}[1]{}
\newcommand*{\Degree}[1]{}\newcommand*{\EMail}[1]{}
\newcommand*{\AuthorAddress}[1]{}
\newcommand{\AuthorInfo}[7]{%
\AuthorID{#1}\LastName{#2}\FirstName{#3}\ShortFirstName{#4}%
\Degree{#5}\EMail{#6}\AuthorAddress{#7}}
\newcommand{\ud}{\mathrm{d}}
\begin{document}
\AuthorInfo{1}
    {Zeitlin}
    {Michael}
    {M.}
    {Dr.}
    {zeitlin@math.ipme.ru}
    {Russian Academy of Sciences, Institute of Problems of Mechanical Engineering,                    
     Mathematical Methods in Mechanics Group, V.O., Bolshoj pr., 61, 199178, 
     St.~Petersburg, Russia}
\AuthorInfo{2}
    {Fedorova}
    {Antonina}
    {A.}
    {}
    {anton@math.ipme.ru}
    {Russian Academy of Sciences, Institute of Problems of Mechanical Engineering,                    
     Mathematical Methods in Mechanics Group, V.O., Bolshoj pr., 61, 199178, 
     St.~Petersburg, Russia}
\Symposium{0}

\Session{15} 
\makeheadings \markboth{\jtitle}{\symptitle}
\begin{minipage}[t]{180mm}
\thispagestyle{empty}
\vspace{20mm}

\begin{center}
{\Large\bf Nonlinear Motion in Electromagnetic}

\vspace{7mm}

{\Large\bf Fields via Multiscale Expansions} 

\vspace{20mm}

{\large\bf Antonina N. Fedorova, Michael G. Zeitlin}

\vspace{20mm}

Mathematical Methods in Mechanics Group \\

Institute of Problems of Mechanical Engineering (IPME RAS)\\ 

Russian Academy of Sciences \\

Russia, 199178, St. Petersburg, V.O., Bolshoj pr., 61\\

zeitlin@math.ipme.ru, anton@math.ipme.ru\\

http://www.ipme.ru/zeitlin.html\\

http://www.ipme.nw.ru/zeitlin.html

\vspace{20mm}
{\bf Abstract}

\vspace{10mm}

\begin{tabular}{p{100mm}}
The consideration of dynamics of relativistic beams/particles 
is based on variational approach to rational (in dynamical     
variables) approximation for equations of motions. It allows to           
control contribution from each scale of underlying multiscales                 
and represent solutions via exact nonlinear eigenmodes expansions.        
Our approach is                                                                    
based on methods provided possibility to work with well-localized bases        
in phase space and good convergence properties of the corresponding                
expansions.

\vspace{10mm}

Presented: GAMM Meeting, February, 2001, ETH, Z\"urich

\vspace{5mm}

Published: PAMM, Volume 1, Issue 1, pp. 432-433, Wiley-VCH, 2002

\end{tabular}

\end{center}
\end{minipage}
\newpage

\vspace*{15mm}\hspace*{18mm}
\begin{minipage}[t]{157mm}


\Author{Fedorova, A.; Zeitlin M.}
\vspace*{0.4cm}

\Title{Nonlinear Motion in Electromagnetic Fields via Multiscale Expansions}
\end{minipage}
\vspace*{3.5mm}


\begin{Abstract}%
The consideration of dynamics of relativistic beams/particles 
is based on variational approach to rational (in dynamical     
variables) approximation for equations of motions. It allows to           
control contribution from each scale of underlying multiscales                 
and represent solutions via exact nonlinear eigenmodes expansions.        
Our approach is                                                                    
based on methods provided possibility to work with well-localized bases        
in phase space and good convergence properties of the corresponding                
expansions.                    
\end{Abstract}

We consider as the main example the particle motion in
storage rings in standard approach based on
consideration in [1].
Starting from Hamiltonian, which describes classical dynamics in
storage rings
$
{\cal H}(\vec{r},\vec{P},t)=c\{\pi^2+m_0^2c^2\}^{1/2}+e\phi
$
and using Serret--Frenet parametrization, we have after
standard manipulations with truncation of power series expansion of
square root the 
rational approximation (up to n-poles) for Hamiltonian of orbital motion
in machine coordinates.
So, our problems may be formulated as the systems of ordinary differential            
equations                                                               
\begin{eqnarray}\label{eq:pol0}                                
Q_i(x)\frac{\ud x_i}{\ud t}=P_i(x,t),\quad x=(x_1,..., x_n),\quad
i=1,...,n, \quad                                                                        
 \max_i  deg \ P_i=p, \quad \max_i deg \  Q_i=q                  
\end{eqnarray}                                                 
with initial (or boundary) conditions $x_i(0)$, $x_i(T)$ and  $P_i,  Q_i$ are not more    
than polynomial functions of dynamical variables $x_j$                                 
and  have arbitrary dependence on time. 
Of course, we consider such $Q_i(x)$ which do not lead to the singular
problem with $Q_i(x)$, when $t=0$ or $t=T$, i.e. $Q_i(x(0)), Q_i(x(T))\neq\infty$, 0.
We'll consider these  equations as the following operator equation.
Let $L$ be an arbitrary nonlinear (rational) matrix differential operator of the first order with matrix dimension d
corresponding to the system of equations (1), 
which acts on some set of functions
$\Psi\equiv\Psi(t)=\Big(\Psi^1(t),\dots,\Psi^d(t)\Big), \quad t \in\Omega\subset R$
from $L^2(\Omega)$:
\begin{equation}
L\Psi\equiv L(R,t)\Psi(t)=0,\qquad  R\equiv R(t,\partial /\partial t, \Psi).
\end{equation}
Let us consider now the N mode approximation for solution as the following ansatz (in the same way
we may consider different ansatzes):
\begin{equation}
\Psi^N(t)=\sum^N_{r=1}a^N_{r}\psi_r(t)
\end{equation}
We shall determine the coefficients of expansion from the following variational conditions
(different related variational approaches are considered in [1]-[4]):
\begin{equation}
L^N_{k}\equiv\int(L\Psi^N)\psi_k(t)\ud t=0
\end{equation}
We have exactly $dN$ algebraical equations for  $dN$ unknowns $a_{r}$.
So, variational approach reduced the initial problem (1) to the problem of solution 
of functional equations at the first stage and some algebraical problems at the second
stage. 
Here $\psi_k(t)$ are useful basis functions of  some functional
space ($L^2, L^p$, Sobolev, etc) corresponding to concrete
problem. 
As result we have the following reduced algebraical system
of equations (RSAE) on the set of unknown coefficients $a_i^N$ of
expansions (3):
\begin{eqnarray}\label{eq:pol2}
L(Q_{ij},a_i^N,\alpha_I)=M(P_{ij},a_i^N,\beta_J),
\end{eqnarray}
where operators L and M are algebraization of RHS and LHS of initial problem
(\ref{eq:pol0}).
$Q_{ij}$ are the coefficients (with possible time dependence) of LHS of initial
system of differential equations (\ref{eq:pol0}) and as consequence are coefficients
of RSAE.
 $P_{ij}$ are the coefficients (with possible time dependence) of RHS
of initial system of differential equations (\ref{eq:pol0}) and as consequence
are the coefficients of RSAE.
$I=(i_1,...,i_{q+2})$, $ J=(j_1,...,j_{p+1})$ are multiindexes, by which are
labelled $\alpha_I$ and $\beta_I$,  the other coefficients of RSAE (\ref{eq:pol2}):
\begin{equation}\label{eq:beta}
\beta_J=\{\beta_{j_1...j_{p+1}}\}=\int\prod_{1\leq j_k\leq p+1}\psi_{j_k},\qquad
\alpha_I=\{\alpha_{i_1}...\alpha_{i_{q+2}}\}=\sum_{i_1,...,i_{q+2}}\int
\psi_{i_1}...\dot{\psi_{i_s}}...\psi_{i_{q+2}},
\end{equation}
where p (q) is the degree of polynomial operator P(Q) (\ref{eq:pol0}),
$i_\ell=(1,...,q+2)$, $\dot{\psi_{i_s}}=\ud\psi_{i_s}/\ud t$.
According to [1]-[4] we may extend our approach to the case when we have additional
constraints on the set of our dynamical variables $\Psi$ or $x$.
In this case by using the method of Lagrangian multipliers we again may apply the same approach but
for the extended set of variables. As result we receive the expanded system of algebraical equations
analogous to the system (5). Then, after reduction we again can extract from its solution the coefficients 
of expansion (3).  
Now, when we solve RSAE (\ref{eq:pol2}) and determine
unknown coefficients from formal expansion (3) we therefore
obtain the solution of our initial problem.
It should be noted if we consider only truncated expansion (3) with N terms
then we have from (\ref{eq:pol2}) the system of $N\times d$ algebraical equations
with degree $\ell=max\{p,q\}$
and the degree of this algebraical system coincides
with the degree of initial differential system.
So, we have the solution of the initial nonlinear
(rational) problem  in the form
\begin{eqnarray}\label{eq:pol3}
x(t)=x(0)+\sum_{k=1}^Na_k^N \psi_k(t),
\end{eqnarray}
where coefficients $a_k^N$ are roots of the corresponding
reduced algebraical (polynomial) problem RSAE (\ref{eq:pol2}).
Consequently, we have a parametrization of solution of initial problem
by solution of reduced algebraical problem (\ref{eq:pol2}).
The problem of
computations of coefficients $\alpha_I$ , $\beta_J$
(\ref{eq:beta}) of reduced algebraical
system
may be explicitly solved in wavelet approach.
The obtained solutions are given
in the form (\ref{eq:pol3}),
where
$\psi_k(t)$ are wavelet basis functions. In our case $\psi_k(t)$
are obtained via multiresolution expansions and represented by
compactly supported wavelets. 
Because affine
group of translation and dilations is inside the approach, this
method resembles the action of a microscope. We have contribution to
final result from each scale of resolution from the whole
infinite scale of spaces:
$
...V_{-2}\subset V_{-1}\subset V_0\subset V_{1}\subset V_{2}\subset ...,
$
where the closed subspace
$V_j (j\in {\bf Z})$ corresponds to  level j of resolution, or to scale j.
This multiresolution functional space decomposition corresponds to exact nonlinear
eigenmode decompositions (7).
It should be noted that such representations 
give the best possible localization
properties in the corresponding (phase)space/time coordinates. 
In contrast with different approaches formulae (7)  do not use perturbation
technique or linearization procedures 
and represent dynamics via generalized nonlinear localized eigenmodes expansion.  
So, by using wavelet bases with their good (phase)space/time      
localization properties we can construct high-localized (coherent)  structures in      
nonlinear systems with collective/complex behaviour.
As a result our N mode construction (7) gives the following representation for solution of equations (1):
\begin{eqnarray}
x(t)&=&x_{N}^{slow}(t)+\sum_{i\ge N}x^i(\omega_it),
\quad \omega_i \sim 2^i
\end{eqnarray}
where $x^r(t)$ may be represented by some family of (nonlinear)
eigenmodes and gives as a result the multiresolution/multiscale representation in the
high-localized wavelet bases.

\begin{Acknowledgements}
We would like to thank ETH, Zurich for hospitality and support, 
which gave us the possibility to present our two papers during
GAMM 2001 Annual Meeting in Zurich and Prof. Martin Gutknecht
for permanent help and encouragement.
\end{Acknowledgements}

\begin{References}

\item {\normalsize \sc Fedorova, A., Zeitlin M.}:
Symmetry, Hamiltonian Problems and Wavelets in Accelerator Physics,
American Institute of Physics, CP, {\bf 468}, 
Nonlinear and Collective Phenomena in Beam Physics 
(1999), 69--93.

\item {\normalsize \sc Fedorova, A., Zeitlin M.}:
Wavelets in Optimization and Approximations; 
Math. and Comp. in Simulation, {\bf 46} (1998), 527--534.

\item  {\normalsize \sc Fedorova, A., Zeitlin M.}:
Variational-Wavelet Approach to RMS Envelope Equations;
The Physics 
of High Brightness Beams, World Scientific (2000), 235--254.

\item {\normalsize \sc Fedorova, A.; Zeitlin M..}: 
Quasiclassical Calculations for Wigner Functions via 
Multiresolution, 
Quantum Aspects of Beam Physics, World 
Scientific (2001); Los Alamos preprint, physics/0101006.

\end{References}

\begin{Address}
{\sc Dr. Michael Zeitlin,}
Russian Academy of Sciences,                                                
Institute of Problems of Mechanical Engineering, \\                             
V.O., Bolshoj pr., 61, 199178, St.~Petersburg, Russia \\
email: zeitlin@math.ipme.ru, 
http://www.ipme.ru/zeitlin.html, http://www.ipme.nw.ru/zeitlin.html

\end{Address}

\end{document}